\documentstyle[twoside,fleqn,espcrc2,graphicx]{article}

\title{s+d pairing in orthorhombic phase of copper-oxides}

\author{N.M. Plakida   and V.S.~Oudovenko\\[0.3cm]
{Joint Institute for Nuclear Research, 141980 Dubna, Russia}      }

\begin{document}

\begin{abstract}
A microscopical theory of electronic  spectrum and  superconductivity
is formulated within the two-dimensional anisotropic  $t-J$ model
with $t_{x}\neq t_{y}$ and $J_{x}\neq J_{y}$. Renormalization of
electronic  spectrum   and  superconductivity mediated by
spin-fluctuations are investigated within the Eliashberg equation
in the weak coupling approximation. The gap function has $d+s$
symmetry with the extended $s$-wave component
being proportional to the asymmetry $ t_{y} - t_{x}$.
Some experimental consequences of the obtained results  are  discussed.
\vspace{1pc}
\end{abstract}

\maketitle

Recently the $d$-wave symmetry of superconducting pairing in cuprates
was unambiguously confirmed by observation of half-integer magnetic
flux quantum~\cite{kirt}. In a tetragonal phase of copper oxides the
$s$-wave component must be  strongly suppressed  due to  on-site Coulomb
correlations.  For the  2D $t$-$J$ model it follows from  the constraint
of no double occupancy on a single site given by the identity~\cite{plak89}:
\begin{equation}
\langle \hat{a}_{i, \sigma} \hat{a}_{i,-\sigma}\rangle =
\frac{1}{N} \sum_{k_{x}, k_{y}}
\langle \hat{a}_{{\bf k}, \sigma} \hat{a}_{-{\bf k}, -\sigma}\rangle = 0 .
\label{1}
\end{equation}
for the projected  electron operators
$\hat{a}_{i,\sigma}=a_{i,\sigma}(1-n_{i,-\sigma})$.
The anomalous correlation function
$\langle \hat{a}_{k, \sigma} \hat{a}_{-k, -\sigma}\rangle$
is proportional to the gap function, $\Delta(k_{x},k_{y})$,
multiplied by a positive function  symmetric in respect to the $D_{4h}$
point group which defines the symmetry of the Fermi surface (FS). Therefore
to satisfy the condition~(\ref{1}) the gap function should have a lower
symmetry, e.g. $B_{1g}$, "$d$-wave"  symmetry (see, e.g.~\cite{rice87}):
$
\Delta_{d}(k_{x},k_{y})=  - \Delta_{d}(k_{y},k_{x}).
$

In the orthorhombic phase the FS has a lower symmetry,
e.g., $D_{2h}$, and  the condition~(\ref{1}) can be fulfilled for a gap
function of the same symmetry (within  the   $E_{1g}$ irreducible
representation) which  can be written in a general form  ("$d+s$"):
\begin{equation}
\Delta(k_{x},k_{y})= \Delta_{d}(k_{x},k_{y}) +
\epsilon \; \Delta_{s}(k_{x},k_{y}).
\label{2}
\end{equation}
where $\Delta_{s}(k_{x},k_{y})=\Delta_{s}(k_{y},k_{x})$ is "the extended
$s$-wave" component.

In the present paper we calculate superconducting $T_c$ for the
2D $t$-$t'$-$J$ model within the theory developed by us in~\cite{plak99},
both in  tetragonal  and  orthorhombic phases.
The orthorhombic ($D_{2h}$) distortion is taken into account
by introducing the asymmetric hopping parameters $t_{ij}$ and the exchange
interaction $J_{ij}$ for the nearest neighbors (n.n.) in the form:
$t_{x/y} = t (1 \pm \alpha)$, $J_{x/y} = J(1 \pm \beta) $
where the asymmetry parameters are supposed to be  small quantities:
$ \alpha \sim \beta \sim 0.1$. For the next n.n.,  $t_{ij} =t'.$

The Dyson equations for the matrix Green function (GF) in the Nambu
notation was obtained by the  equation of motion method for the Hubbard
operators as described in~\cite{plak99}. For estimation of the role of
orthorhombic deformation we consider here only the  weak-coupling
approximation.  However, to take into account strong electronic correlations
in the $t$-$J$ model due to restricted hopping in the singly occupied
subband we write  the single-electron spectral density in the form:
$\;A(k,\omega) \simeq Z_{k}\delta(\omega+{\mu} -
E_{k}) + A_{inc}(\omega)$.
The quasiparticle weight $ Z_{k}$ and  the incoherent part $ A_{inc}(\omega)$
are coupled  by the sum rule for the spectral density:
$
\int_{-\infty }^{+\infty }d\omega
A(k, \omega)  = 1 - {n}/{2} .
$
To fix the value of $Z_{k}$  we assume that  the FS
for quasiparticles with  spectrum $E_{k}$  obeys the
Luttinger theorem: the average number of electrons is equal to the
number of states in $k$-space  below the chemical potential $ \mu$:
$ \;n = (1/N) \sum_{k, \sigma} \{\exp [E_{k} - \mu)/ T] + 1\} ^{-1}  .
$
>From these conditions we have  estimations:
$Z \simeq (1-n)/ (1-n/2)$ and
$ A_{inc}(\omega) \simeq (n/2)^{2}/(1-n/2)(W-\Gamma)$
where we have suggested that  the coherent band lies in the range
$-\Gamma \leq \omega \leq \Gamma$  while  the incoherent band lies
below the coherent band in the range
$-W \leq \omega \leq -\Gamma$.
By taking into account the renormalization of the coherent part of
the spectral weight by $Z_{k}$  we  write the
equation for the gap  in the weak coupling approximation in the form:
\begin{equation}
\Delta_{k} =  \frac{1}{N} \sum_{q} K(q,k-q)
\frac{Z_{q}^{2}\; \Delta_{q}}{2 \Omega_{q}}
\tanh\frac{\Omega_{q}}{2T}.
\label{3}
\end{equation}
where $\Omega_{q}= [(E_{q}-\mu)^2 + \Delta_{q}^2]^{1/2}$ and
$E_{q} \simeq  - t_{eff}[\gamma(q)+\alpha \eta(q)]-t^{'}_{eff}\gamma'(k)$
with renormalized due to strong correlations hopping parameters
and  $ \gamma(q)=(1/2)(\cos {q}_{x}+\cos{q}_{y}),\;
\eta(q)=(1/2)(\cos{q}_{x}-\cos{q}_{y}),
\gamma'(q)=\cos{q}_{x}\cos{q}_{y}$~\cite{plak99}.
$K(q,k-q) = \{2g(q,k-q)
- \lambda(q,k-q) \}$ with the vertex $g(q,k-q)= t(q)-J(k-q)/2$
and  $\lambda(q,k-q)= g{^2} (q,k-q) \chi(k-q)$. The first term in the vertex,
$t(q)$, is due to the kinematical interaction caused by constraints
and the second one, $J(q)$, is the exchange coupling. They have different
$q$-dependence and are effective at different doping.
The spin-fluctuation coupling in $\lambda(q,k-q)$ is defined by the static
spin susceptibility $\chi(k-q)$ for which  we used the model
$\chi(q) = {\chi_0} / [1+ \xi^2 (1+ \gamma(q))] $
where the antiferromagnetic (AFM) correlation  length $\xi$ is a
fitting parameter while  $\chi_0$ is normalized by the condition:
$ {1}/{N}\sum_{i} \langle {\bf S}_{i}{\bf S}_{i} \rangle =
   ({3}/{4})n $.

We performed numerical solution of Eq.~(\ref{3}) for the gap in the
form~(\ref{2}) with
$\Delta_{d}(k_{x},k_{y})= \Delta \eta(k)$
and
$\Delta_{s}(k_{x},k_{y})= \Delta \gamma(k)$.
By taking into account the
constraint of no double occupancy, Eq.(\ref{1}),
we estimate the weight $\epsilon$ of the $s$-component.
The critical temperature $T_{c}(\delta)$ (in units of $t$)
is  shown on Fig.1 in the tetragonal, $\alpha=0.0$,
(bold line) and orthorhombic,  $\alpha=0.1$, (dashed line)  phases for
$J=0.4t,\; \xi=2, \;t'=0.0$. Suppression of $T_{c}$ in the  orthorhombic
phase is due to a deformation of the FS resulting in a less favourable
electron  pairing  by the AFM spin fluctuations.
Increasing of AFM interaction due to larger $J$ or/and  $\xi$ strongly
enhances $T_c$ though does not change the shape of the curve.
Its maximum at $\delta \simeq 0.35$ is due to an interplay between the
shape of the FS (defined by the quasiparticle spectrum $E_{q}$)
and the coherent spectral weight $Z^2$ in Eq.(\ref{3}). Particularly,
for $t'/t = -0.1\; (+0.1)$ we observed a strong enhancement (suppression)
of $T_{c}(\delta)$ due to change of the FS.  In the orthorhombic phase
the $s$-wave component with  $\epsilon \leq -1$ (depending on the doping)
appears in  Eq.(\ref{2}) shifting  4 nodes of the gap at the FS  from the
diagonals $k_{x}=\pm k_{y}$ in a tetragonal phase as in
Ref.~\cite{donovan95}.
\mbox{}\\[-1.0cm]
\begin{figure}[htb]
\includegraphics[width=7cm]{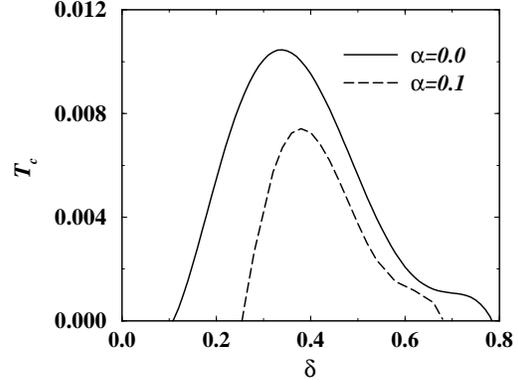}\\[-1.5cm]
\caption{$T_{c}(\delta)$
in the tetragonal  (bold line) and orthorhombic (dashed line)  phases
($t'=0.0$).}
\label{fig:1}
\end{figure}
\mbox{}\\[-1.0cm]
To conclude, in the present paper we estimate the role of orthorhombic
deformation in cuprates within the $t$-$t'$-$J$ model  by solving the
equation (\ref{3}) for a gap of general symmetry (\ref{2}) with constraint
of no double occupancy, Eq.(\ref{1}).
The obtained dependence $T_{c}(\delta)$ in Fig.1
can explain  a suppression of $T_c$ in the orthorhombic phase
that observed  in LSCO~\cite{goko99} and
anisotropic pressure dependence of $T_c$ in YBCO~\cite{welp92}.
Quite a large "$s$"-wave component in the gap (\ref{2})is  in accord
with a nonzero tunnelling along the $c$-axis  in YBCO~\cite{kouznets97}.

\end{document}